\documentclass[conference]{IEEEtran}
\IEEEoverridecommandlockouts
\usepackage{cite}
\usepackage{amsmath,amssymb,amsfonts}
\usepackage{algorithmic}
\usepackage{graphicx}
\usepackage{textcomp}
\usepackage{xcolor}
\usepackage{algorithmic}
\usepackage[ruled,vlined,linesnumbered,noresetcount]{algorithm2e}
\usepackage{amsmath}
\usepackage{listings}
\usepackage{verbatim} 
\usepackage{color,soul}
\usepackage{booktabs}
\usepackage{comment}
\usepackage{algorithmic}
\usepackage[ruled,vlined,linesnumbered,noresetcount]{algorithm2e}
\usepackage{listings}
\usepackage{verbatim} 
\usepackage{balance}
\usepackage{caption}
\usepackage{subcaption}
\usepackage{xcolor}
\usepackage{geometry}
\newcommand{\HA}[1]{{\color{black}{#1}}}
\newcommand{\HAR}[1]{{\color{black}{#1}}}
\newcommand{\TD}[1]{{\color{purple}{#1}}}

\def\BibTeX{{\rm B\kern-.05em{\sc i\kern-.025em b}\kern-.08em
    T\kern-.1667em\lower.7ex\hbox{E}\kern-.125emX}}
\begin{document}

\title{Latency and Privacy-Aware Resource Allocation in Vehicular Edge Computing\\

}

\author{\IEEEauthorblockN{ Hossein Ahmadvand}
\IEEEauthorblockA{
\textit{}\\
}

\and
\IEEEauthorblockN{Fouzhan Foroutan}
\textit{}


}

\maketitle

\thispagestyle{plain}
\pagestyle{plain}
\begin{abstract}
The rapid increase in the number of connected vehicles has led to the generation of vast amounts of data. As a significant portion of this data pertains to vehicle-to-vehicle and vehicle-to-infrastructure communications, it is predominantly generated at the edge. Considering the enormous volume of data, real-time applications, and privacy concerns, it is crucial to process the data at the edge. Neglecting the management of processing resources in vehicular edge computing (VEC) could lead to numerous challenges as a substantial number of vehicles with diverse safety, economic, and entertainment applications, along with their data processing, emerge in the near future  \cite{sedar2023comprehensive}.
Previous research in VEC resource allocation has primarily focused on issues such as response time and privacy preservation techniques. However, an approach that takes into account privacy-aware resource allocation based on vehicular network architecture and application requirements has not yet been proposed. In this paper, we present a privacy and latency-aware approach for allocating processing resources at the edge of the vehicular network, considering the specific requirements of different applications. Our approach involves categorizing vehicular network applications based on their processing accuracy, real-time processing needs, and privacy preservation requirements. We further divide the vehicular network edge into two parts: the user layer (OBUs) is considered for processing applications with privacy requirements, while the allocation of resources in the RSUs and cloud layer is based on the specific needs of different applications.
In this study, we evaluate the quality of service based on parameters such as privacy preservation, processing cost, meeting deadlines, and result quality. Comparative analyses demonstrate that our approach enhances service quality by 55\% compared to existing state-of-the-art methods.

\end{abstract}

\section{Introduction}

\HA{ 

As outlined in the 2050 roadmap established by ERTRAC, the future envisions a comprehensive network of interconnected vehicles enabled by real-time communication \cite{ERTRAC}. This trajectory paves the way for the proliferation of large-scale self-driving vehicles in the years to come, thereby presenting a multitude of challenges in the domains of processing infrastructure, privacy, and latency management within vehicular networks.
As vehicular networks expand rapidly and extensively, addressing the specific needs of this technology has become essential. Real-time responsiveness and low-latency processing are critical for the efficient operation of these networks.



A significant portion of the data generated in vehicular networks stems from vehicle-to-vehicle (V2V) and vehicle-to-infrastructure (V2I) interactions (Figure \ref{fig:Vehicular-Fog-Architecture}). Consequently, much of this data is processed at the edge or near its point of generation.

The vehicular edge computing (VEC) environment encompasses a diverse range of heterogeneous devices, characterized by limited and heterogeneous edge processing infrastructure \cite{sedar2023comprehensive}. 

Consequently, much of this data is processed at the edge or near its point of generation \cite{he2017privacy}.

To address this challenge, our study introduces a comprehensive approach that considers priorities such as application type, acceptable latency, required accuracy, desired privacy level, and whether the application data is centralized or distributed.

Considering these factors, we determine the optimal processing location and method for each application.
The works in \cite{zheng2020blockchain, xu2021blockchain, guo2020blockchain, dargahi2021integration} have addressed the privacy-preserving and edge processing using blockchain. The problem of privacy preservation is a very important issue that we also address in this article. However, privacy preservation cannot be achieved without considering the allocation of resources, regardless of the type of application and its requirements.
The authors in \cite{lu2021accelerating, fan2020secure, brousmiche2018hybrid, zhang2019data, kong2020achieving, zheng2015heterogeneous} have considered data sharing and data storage problem of vehicular edge computing. Simultaneous attention to privacy and latency management is an issue that these researchers have not addressed. While in this research, we have first identified the processing elements in the processing edge by examining the architecture of the vehicular network. Then, with our attention to latency management and privacy preservation, we have provided a solution for processing resource allocation.
\HAR{Similar work has been done in Mobile Edge Computing (MEC). Researchers have presented a model for measuring the quality of service and privacy in terms of processing resources in MEC
\cite{he2017privacy} and \cite{min2018learning}. In this article, we have looked at this issue in vehicular networks from a different perspective. Also, the model presented by us is different from their model according to the characteristics of the vehicular networks.}
In this paper, we discuss edge processing in vehicular networks and their requirements. We categorize the applications and identify the requirements of each type of application. These requirements include real-time processing, privacy, and processing accuracy. According to these parameters, we decide on the data processing in vehicular networks.  We have investigated the architecture of the vehicular networks to identify the processing elements at the edge of the vehicular networks. 
Figure \ref{fig:Vehicular-Fog-Architecture} shows our reference architecture of vehicular edge computing. We consider OBUs and RSUs as processing elements at the edge of the network in this paper. Also, in this architecture, we have specified the user layer. This layer consists of the vehicles' OBUs. For applications that need a high level of privacy, we use the user layer (which is part of the edge layer). In this way, the information of each user in those applications is processed only on the user's device. 

\begin{figure}[htbp]
\centerline{\includegraphics[width=\columnwidth]{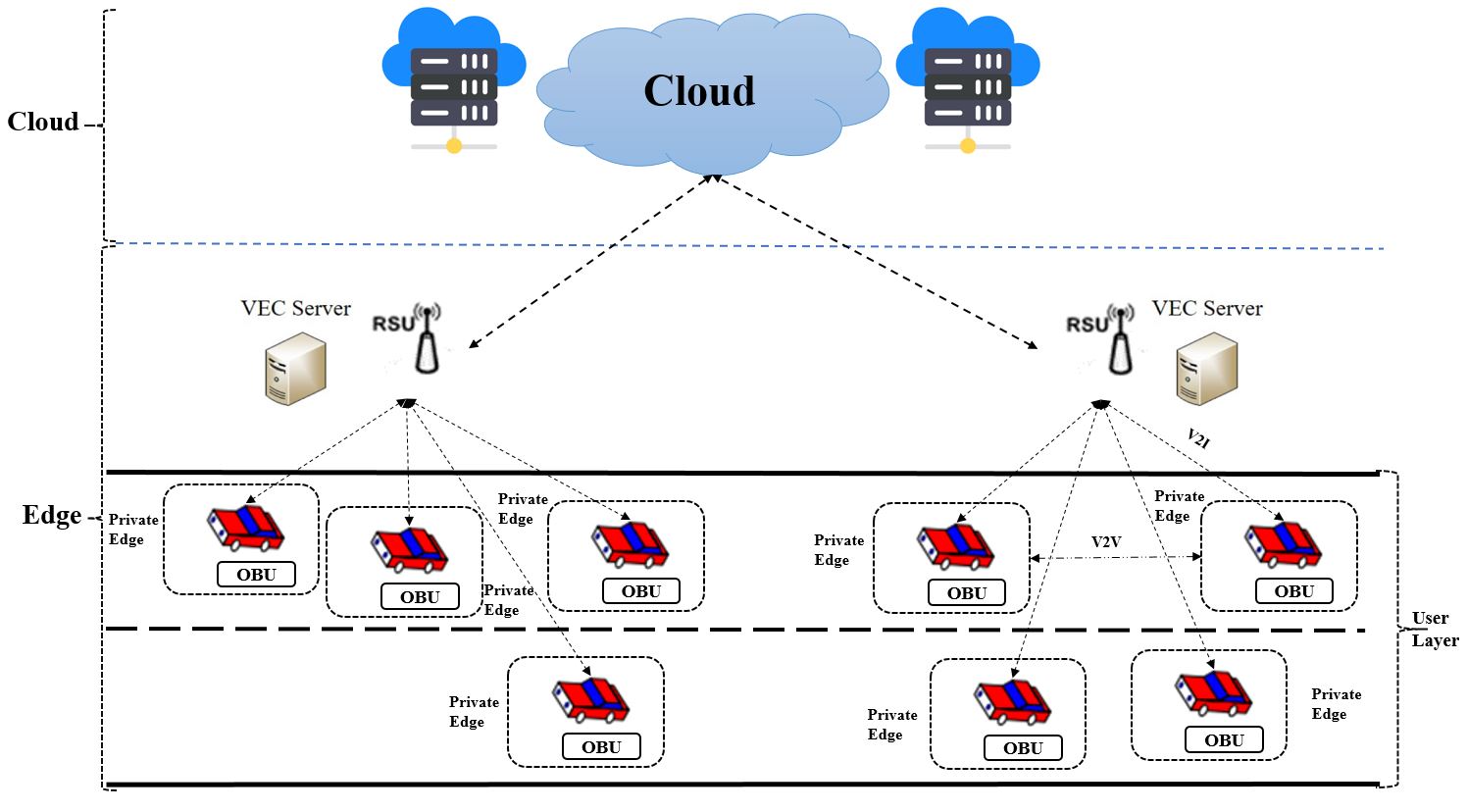}}
\caption{Vehicular Edge Architecture \cite{hoque2019towards}}
\label{fig:Vehicular-Fog-Architecture}
\end{figure}

\textbf{Our Contribution.} In this study, we will present the following contributions:(1) \HAR{In this research, we consider heterogeneity in the processing infrastructure at the VEC. For this purpose, we have examined the architecture of the vehicular network identified its various processing components, and examined them for use in our approach. } (2) In this study, we consider privacy in providing solutions. based on this, we determine a layer in the edge that processes the application with privacy preservation (User Layer). (3) In addition to managing heterogeneous resources in the edge processing environment and privacy-preserving, in this study, we also consider the tolerable latency of each application. based on this, we suggest a suitable resource and location for processing based on application type. 

In this paper, we discuss the implementation of a recommender system in the field of privacy and edge processing. In this recommender system, according to the requirements of different applications (the amount of tolerable delay, the amount of data privacy, and the amount of infrastructure required for processing), the data processing location is suggested to the user. This proposal can be cloud, edge, or a combination of both of them.}

The rest of the article is as follows:
First, in Section II, we talk about the existing works. In Section III, we describe the problem formulation. We examined our approach with the experiments in Section V. In section VI, we summarize the issues raised in the article and presented existing directions for future research.

\section{Related Works}

\HA{
In this section, we review previous works in the field of heterogeneity in edge processing. We have categorized the previous works into two categories. These categories are: Resource allocation in VEC and Privacy in VEC. 

\textbf{Resources allocation in VEC.} 
In this article ~\cite{zhang2019heteroedge}, the authors examine the use of edge processing capacity to process social network data. They have considered important issues such as limitation of processing capacity, real-time processing, and security concerns in such kind of processing. Similar to us the authors in \cite{sookhak2017fog} consider reducing latency and limitation-aware resource allocation. they also considered the various quality of service and various applications' requirements. 
One of the important issues that we have addressed in this paper is the offloading of the workload of the vehicular network. In  article~\cite{liu2022mobility}, d the off-loading problem in vehicular edge computing is considered. For this, the researchers have considered the processing and communication time as optimization parameters.
The authors in \cite{xiao2021vehicle} have presented a novel federated learning approach by using onboard units as a processing resource. We also have an idea like theirs, but in our work, other requirements also have been considered. Privacy preservation and priority of applications and users are among the requirements that we have considered in our model. They also consider multi-hob between vehicles for task offloading. We didn't consider offloading between vehicles due to privacy concerns. 
 The authors in \cite{zhu2021deep} formulate the competitive interactions between the VEC servers and vehicles as a two-stage Stackelberg game with the VEC servers as the leading players and the vehicles as the followers. 
The study \cite{lakhan2021mobility} proposes local search-aware task scheduling and mobility-aware offloading schemes based on linear programming. The study devises locally Based Task Scheduling (LSBTS), Mobility Aware Blockchain-Enabled
offloading scheme (MABOS) scheme to optimize functions for applications.
Various previous research works have investigated vehicular edge computing in \cite{meneguette2021vehicular}. None of the previous works have discussed the subject from our point of view.
 We consider the quality of service in vehicular edge computing in our model. The quality of service is also considered in\cite{zhu2022computation}
 The objective of this research is to maximize the long-term management profit of the network operator. We also consider the QoS in our work, but our definition of QoS is very different and includes processing cost and privacy. The authors in \cite{ahmadvand2021big} have considered the impact of data skew on edge cloud computing. In their study, they considered the data skew and the limitation of processing resources at the edge. For this reason, they have managed to use the processing resources available at the edge in a better way. In \cite{raza2020efficient}, the authors study the partial task offloading problem in vehicular edge computing in an urban scenario. Where the vehicle computes some parts of a task locally and offloads the remaining task to a nearby vehicle and to the Vehicular Edge Computing server subject to the maximum tolerable delay and vehicle’s stay time.
This paper \cite{ren2021blockchain} introduces the VEC network architecture controlled by a two-layer distributed SDN as well as blockchain technology into the architecture as the carrier of network topology information sharing. Vehicular service execution delay (edge), system energy consumption, and the blockchain system throughput are jointly considered to formulate a problem optimization model.
This paper \cite{guo2020blockchain} proposes that the free resource displayed on non-mining devices and edge cloud can be selected to construct a collaborative mining network (CMN) to execute mining tasks for mobile blockchain. In the CMN, mobile users decide whether to offload mining tasks to sharing devices in the CMN or edge cloud. The authors have studied the impact of various configurable parameters on the performance of CMNs.
\HAR{In this article \cite{xia2022privacy}, the authors have presented an incentive approach to increase the participation of vehicles in processing and balancing the workload. They do not mention the requirements of different applications in terms of privacy and do not provide a solution for privacy-preserving.}
The authors in \cite{lu2021accelerating, fan2020secure, brousmiche2018hybrid, zhang2019data, kong2020achieving, zheng2015heterogeneous} have considered data sharing and data storage problem of vehicular edge computing.


 \textbf{Privacy in VEC.}
The authors in \cite{zheng2020blockchain}, combine blockchain and deep reinforcement learning for the edge cloud offloading system in the VANET network and jointly investigate access control and computation offloading. They have increased the computational cost and energy consumption in vehicular processing. They have provided an acceptable level of security by using smart contracts. 
The authors in \cite{xu2021blockchain} proposed a blockchain-based secure computation offloading scheme in a vehicular cloud network, which comprises blockchain-based trust management and Deep Reinforcement Learning algorithm-based smart contracts.
The authors in \cite{dargahi2021integration} have analyzed the usages of Blockchain in Connected autonomous vehicles. They have investigated the improvement of privacy levels in various applications by using blockchain CAVs.
The authors in \cite{xue2018fog} proposed a novel edge-to-cloud-based architecture for data sharing in vehicular cloud computing. Their method is a cryptography-based mechanism for data sharing in vehicular environments. The edge and cloud servers are responsible for the complicated computation part with confidentiality and privacy preservation. Edge nodes do not have access to decrypted data which provides a confidentiality level in edge computing. Experiments showed significant performance improvement in response delay reduction and edge devices’ overhead minimization.
\HAR{Similar to our work, the authors in this article \cite{wang2021joint} have provided a solution for offloading the processes in vehicular edge computing. They also consider privacy-preserving in their work and pay attention to it by using a privacy-aware approach for resource allocation. Unlike them, we consider the requirements of various applications in vehicular networks and design our approach based on these requirements. 
In this article, the authors have allocated resources in one. They have used the certificate authority (CA) for privacy-preserving \cite{wang2022copp}. We have not used CA for privacy preservation. Our approach uses only resource allocation for privacy-preserving.
One of the solutions that can deal with the lack of processing resources in the vehicular network is to offload processing from other vehicles \cite{ma2019privacy}. However this issue causes the risk of privacy in the vehicular network. For this reason, we have not considered offloading to other vehicles as part of our approach in this article.

}


Our goal in this paper is to propose an approach for vehicular edge processing by considering various  vehicular applications' requirements. We have considered  the amount of tolerable delay, the required processing infrastructure, and the privacy level of each application in our solution. This issue has not been investigated with considering these parameters and from this point of view in previous works.
}

\section{Problem Statement}
In this section, we discuss the mathematical expression of the problem. In this regard, in the problem formulation section, we discuss the  mathematical expression of objective, limitation, and variables' definitions.
\subsection{Problem Formulation}
The notations that are used in the rest of the paper are presented in  Table~\ref{table1}. 
\begin{table}[htbp]
\centering
	\caption{Notation of our paper}
	\begin{tabular}{p{0.7cm}|p{7cm}}
		\hline
		\textbf{Notation}		       	&			\textbf{Description}   \\ \hline
		
\footnotesize		\textit{E}						& 			Edge \\
  
            \textit{C}						& 			Cloud \\

            \textit{UL} 					& 			User Layer  \\

		\textit{EP} 					& 			Edge Processing \\
		
            \textit{OBU} 					& 			On-Board Unit\\
	    
	      \textit{RSU} 					  & 		 Road side Unit\\
		
		\textit{PT} 					& 			Processing Time \\	
		
		\textit{PC} 					& 			Processing Cost \\

            \textit{SRP}                    &           Server Rent Price\\
				
		\textit{CP} 					& 			Cloud Processing \\
            
            \textit{PA}						& 			Private Application \\

           \textit{PUA}					& 			Public Application \\

            \textit{QoS} 					& 			Quality of Service\\
	    
	    \textit{QoR} 					& 			Quality of Result\\

        \textit{V2V} 					& 			Vehicle to Vehicle  \\
        
        \textit{ACP} 					& 			Accurate Processing \\	

        \textit{RA}						& 			Restricted Application \\

        \textit{RTA} 					& 			Real Time Application  \\

        \textit{V2I}                    &           Vehicle to Infrastructure\\
        
		\textit{AXP} 					& 			Approximate Processing  \\
		
		\textit{SRTA}					& 			Soft Real Time Application \\
            
		\textit{FRTA} 					& 			Firm Real Time Application  \\

  		\textit{HRTA} 					& 			Hard Real Time Application  \\

            \textit{URE} 					& 			Used Resources on the Edge\\

            \textit{URC}                    &           Used Resources on the Cloud\\
            
	    \textit{NMD} 					& 			Number of Missed Deadline\\
	    
		\textit{ACPA} 					& 			Accurate Processing Application  \\

            \textit{ARE} 					& 			Available Resources on the Edge\\

		\textit{AXPA} 					& 			Approximate Processing Application  \\	

            \textit{ARE}                    &           Available Processing Resources on the Edge\\

            \textit{ARE}                    &           Used Processing Resources on the Edge\\

		\hline
		
	\end{tabular}
\label{table1}
\end{table}

\HAR{The objective function is "increasing the Quality of the Service (QoS)". }
We have formulated this objective in Equation~\ref{eq:target}.  We have also considered several constraints in this problem. One of the limitations is the availability of the processing resources at the edge (Equation~\ref{eq:limit}).
\begin{equation}\label{eq:target}
Increase (QoS)\end{equation}
Subject to:
\begin{equation}\label{eq:limit}
ARE>URE
\end{equation}
We have stated the objectives and constraints of the problem in formulas 1-4. Our objectives are to increase the quality of the response, increase the adequacy of the result, and reduce the processing cost.

\subsubsection{Quality of Service}
In this research, we have defined the quality of service as follows. The quality of the service in our work is defined due to the processing cost, number of missed deadlines, and privacy.

\begin{equation}\label{eq:Qos-Eq}
QoS = \frac{1}{Cost+1} * \frac{1}{NMD+1} * Privacy * QoR
\end{equation}

\textbf{Cost.}
In this study, the processing cost is considered only for the cloud part. Therefore, the user will not pay for processing on the edge. The cost of cloud processing is also calculated according to the processing time and the cost of renting the server. We have given the definition of processing cost in Formula 4.

\begin{equation}\label{eq:Privacy-Eq}
PC=  PT*SRP
\end{equation}
\textbf{Number of Missed Deadlines.}
In the above formula, NMD indicates the number of missed deadlines.\\
\textbf{Privacy.}
\HAR{ The most important private data in the car network are location tracking and identity revealing\cite{sedar2023comprehensive}. 
For the definition of privacy, we have used the model presented in this article\cite{he2017privacy}.
 In this research, we have considered the edge (user device) as the private part.  Also, we have specified the privacy of the cloud with a coefficient. Therefore, the criterion of privacy is defined as follows.}

\begin{equation}\label{eq:Privacy-Eq}
Privacy (\%)= \frac{EP(UL)+K(CP)}{EP(UL)+EP(RSUs)+CP} *100
\end{equation}

The K coefficient is related to privacy in cloud processing, which is determined according to the type of used cloud (0-1). We have considered this coefficient equal to 1 here. This means that the considered cloud in this paper provides full privacy preservation.

\textbf{Quality of Result.} 
In some applications, an approximate answer is acceptable as a result. Therefore, the output result is definitely not always equal to the exact answer. The quality of the result is defined as follows.
\begin{equation}\label{eq:Privacy-Eq}
QoR=  ResultAccuracy
\end{equation}

\section{Proposed Approach}\label{sec:approach}

Our approach in this study consciously gives recommendations to the user regarding the privacy of applications, the level of accuracy required, the amount of tolerable delay, and the limitation of processing resources for the location and type.  We consider edge(UL) processing for applications with tight deadlines or applications with private data. We have also divided the edge into two layers. The user layer is used for applications that are private.
We also transfer applications with relaxed deadlines or applications with public data to the cloud for processing or second layer of edge (RSUs)\\
\textbf{Public.} These types of applications are not confidential and can be processed anywhere.\\
\textbf{Restricted.} These types of applications can be processed on the cloud or on the user's device. These applications cannot be processed on RSUs due to their limitations.\\
\textbf{Private.} These types of applications are confidential and should only be processed on the user's device.

We present a heuristic algorithm to solve the problem we defined in the previous section. In this approach, we process private applications and real-time applications on the edge. Applications that are not private and in real-time are processed in the cloud. Also, whether in edge processing or cloud processing, for any application that does not require an accurate answer, approximate processing is used. This is especially important in edge processing.
We also recommend using approximate processing for applications that do not require accurate answers, especially in the case of edge computing due to the lack of processing resources at the edge. 
As shown in Figure \ref{fig:Vehicular-Fog-Architecture} edge processing is suitable for real-time processing, approximate processing, and private data. Also, cloud processing is suitable for accurate processing, soft real-time, and less private data.

\textbf{Algorithm.}
In this part, we will present our algorithm in this paper. The first and second lines of the algorithm show the input and output of the algorithm. We assign real-time and privacy applications to the processing edge (UL). Also, if the processing capacity was left on the edge (UL), we process other approximate processes on the edge (UL). Other applications in this algorithm will be processed on the second layer of the edge (RSUs) or cloud. We consider the second layer of the edge (RSUs) as public processing infrastructure. Also, depending on the type of cloud we decide on the level of its privacy.

\begin{algorithm}\label{algorithm}
{\footnotesize Input: PAs, RAs, PAs, SRTAs, FRTAs, HRTAs \\
footnotesize Output: URE, URC \\
 \While{PAs}  {   
    AXP on  UL for AXPAs   \\
    ACP on UL for ACPAs 

}

\While{RAs}  {
    AXP on  UL for AXPAs \& HRTAs\\
   \eIf{ARE(UL)==1} { AXP on  UL for AXPAs \& FRTAs}
   {
   {AXP on  C for AXPAs \& FRTAs}
   
   }
    AXP on  C for AXPAs \& SRTAs\\

    ACP on  UL for ACPAs \& HRTAs\\
 \eIf{ARE(UL)==1} { ACP on  UL for ACPAs \& FRTAs}
   {
   {AXP on  C for ACPAs \& FRTAs}
   
   }    ACP on  C for ACPAs \& SRTAs\\

}

\While{PUAs}  {
    AXP on  RSUs for AXPAs \& HRTAs\\
    AXP on  RSUs for AXPAs \& FRTAs\\
    AXP on  C for AXPAs \& SRTAs\\

    ACP on  RSUs for ACPAs \& HRTAs\\
    ACP on  RSUs for ACPAs \& FRTAs\\
    ACP on  C for ACPAs \& SRTAs\\

}
}
 \caption{PVEC}
\end{algorithm}

\section{Evaluation}

In this part, we evaluate the work presented in the article. For this purpose, in this section, we introduce the scenarios reviewed in this article, the data sets, and the infrastructures for the evaluations.

\textbf{Scenarios.}The solutions that we have chosen as our competitors are as follows:
\textbf{Linear Search Based Task Scheduling (LSBTS):} This approach uses blockchain to maintain privacy in vehicular edge computing~\cite{lakhan2021mobility}.
\textbf{Random:} This solution allocates resources and is not sensitive to the confidentiality of applications and processing costs. Also, this method does not use approximate calculations.
Our proposed approach is also shown in the diagrams under the title of \textbf{Private Vehicular Edge Computing (PVEC)}.

\textbf{Workloads.} We use known data sets For experiment evaluation, the paper deployed a real scenario of the Every-Where-lab dataset \cite{MARBLE-dataset}. Table \ref{workload} shows the detail of used workloads.

\begin{table}[htbp]
\caption{Workloads)}
\begin{center}
\begin{tabular}{|c|c|c|c|}
\hline

 Workload &  General Tasks & Private Tasks & Real Time Tasks \\
\hline
Healthcare & 1500  & 300 & 200 \\
\hline
E-Transport & 1000  & 150 & 250   \\
\hline
E-Business & 1500  & 250 & 250  \\
\hline

\end{tabular}
\label{workload}
\end{center}
\end{table}

\textbf{Infrastructure.}
 We have considered server configuration according to what has been provided by Amazon EC2 configuration cloud server (memory: 16GB, CPU: 16 Core, cost: 0.959 \$/h)~\cite{Amazon}.
 Also, we have used Raspberry Pi for edge processing infrastructure.
 

\section{Results and Discuss}
We compare our method with other applications in terms of  quality of service,  quality of the result, and processing cost. 

\begin{figure*}
     \centering
     \begin{subfigure}[b]{0.3\textwidth}
         \centering
         \includegraphics[width=\textwidth]{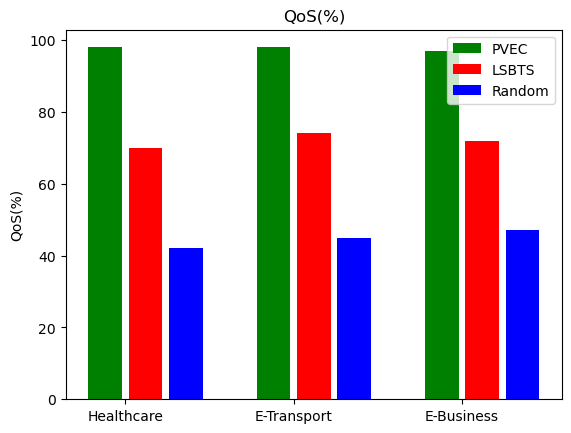}
         \caption{QoS comparison }
         \label{fig:QoS}
     \end{subfigure}
     \hfill
     \begin{subfigure}[b]{0.3\textwidth}
         \centering
         \includegraphics[width=\textwidth]{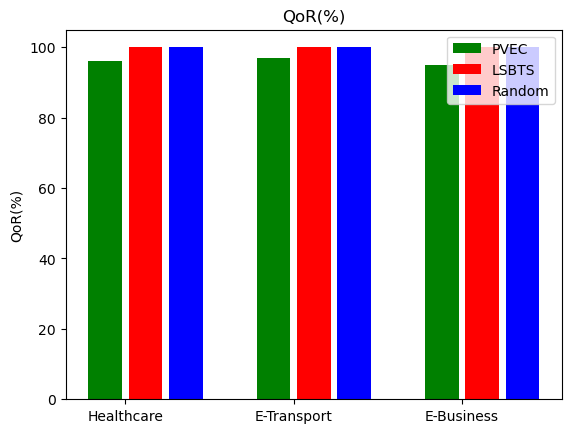}
         \caption{QoR comparison}
         \label{fig:QoR}
     \end{subfigure}
     \hfill
     \begin{subfigure}[b]{0.3\textwidth}
         \centering
         \includegraphics[width=\textwidth]{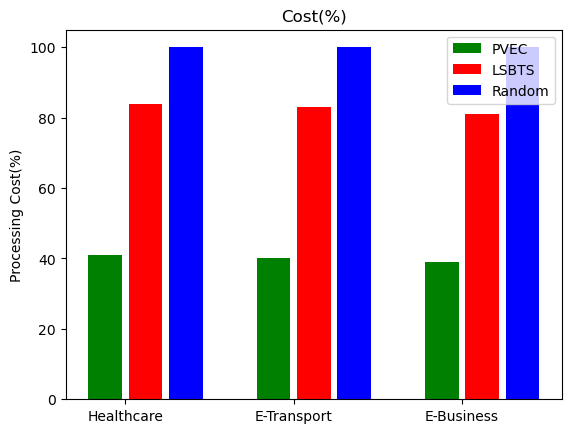}
         \caption{Cost comparison }
         \label{fig:Cost}
     \end{subfigure}
        \caption{The results in different approaches}
        \label{fig:Results}
\end{figure*}

\textbf{Quality of Service.}
The quality of service in our approach is determined based on the processing cost, missed deadlines, and privacy. The exact definition of service quality is given in Formula 3. As shown in Figure \ref{fig:QoS}, our approach increases the quality of the service by 30\%, 25\%, and 24\% compared to LSBTS in the Healthcare, E-Transport, and E-Business workloads.  Compared to the random approach, our approach increases the quality of the service by 55\%, 53\%, and 50\% in the Healthcare, E-Transport, and E-Business workloads.

\textbf{Quality of Result.}
One of our solutions in this article is the use of approximate processing to reduce the consumption of processing resources. Of course, we use approximate processing for applications that do not require accurate answers.
As shown in Figure \ref{fig:QoR}, our approach reduces the quality of the result.  This is because we use approximate calculations for applications that do not require exact answers. In this category of applications, an exact answer is not required and an approximate answer is sufficient. Our approximate processing approach achieves 5\% band error and 95\% confidence interval\cite{ahmadvand2019gapprox}. Therefore, the quality of the result in our approach is reduced to a negligible amount.

\textbf{Cost.}
As mentioned earlier, the processing cost is determined according to the processing time (PT) and the server rental price (SRP). 
As shown in Figure \ref{fig:Cost}, our approach reduces processing cost by 56\%, 49\%, and 53\% in comparison to LSBTS in the Healthcare, E-Transport, and E-Business workloads.  Compared to the random approach, our approach reduces the processing cost by 61\%, 60\%, and 63\% in the Healthcare, E-Transport, and E-Business workloads.

\section{Conclusion and future works}

In this study, we present an approach for private vehicular edge computing.  Our approach considers the deadline, and level of privacy in the allocation step. Our approach used an approximation of the applications that do not require accurate results\cite{ahmadvand2019gapprox, ahmadvand2016using, ahmadvand2019sair}.  Edge processing environment due to the limited processing capacity is suitable for applications that require real-time approximate processing. Using approximate calculations will reduce the usage of processing resources on the edge. We have also divided the edge of the vehicular network into two parts and processed the applications that need privacy on the users' devices (OBUs).
In case of accurate processing, if the application can tolerate the latency of transfer to the cloud, we will process that application on the cloud\cite{ahmadvand2019gapprox}.
Our proposed method has succeeded in increasing the quality of service by 55\% compared to the state of the art.\\
\textbf{Future work}
In the case of complex computation, we will use Federated Learning(FL) and Spiking neural networks(SNNs) for Machine Learning and Deep Learning (ML \& DL). In applying FL, we will look at the heterogeneous resources at the edge as well as privacy preservation. SNNs are also a type of neural network that is compatible with limited processing resources on the edge. In any application, the distribution or centralization of data has a great impact on the choice of processing solution. If the data is centralized, we use the SNN method, and if the data is distributed, we use FL for data processing.

\balance

\bibliographystyle{IEEEtran}
\bibliography{ref.bib}

\end{document}